\documentclass[sigconf]{acmart}

\AtBeginDocument{%
  }

\copyrightyear{2026}
\acmYear{2026}
\setcopyright{cc}
\setcctype{by}
\acmConference[CHI '26]{Proceedings of the 2026 CHI Conference on Human Factors in Computing Systems}{April 13--17, 2026}{Barcelona, Spain}
\acmBooktitle{Proceedings of the 2026 CHI Conference on Human Factors in Computing Systems (CHI '26), April 13--17, 2026, Barcelona, Spain}
\acmPrice{}
\acmDOI{10.1145/3772318.3790670}
\acmISBN{979-8-4007-2278-3/2026/04}

\begin{document}

\title{Understanding User Requirements for Creating Sensor-Powered Smart Car Cabins Through Retrofitting}



\author{Bofan Yu}
\affiliation{%
  \institution{Simon Fraser University}
  \department{School of Computing Science}
  \city{Burnaby}
  \state{British Columbia}
  \country{Canada}
}
\email{bofany@sfu.ca}

\author{BoRui Li}
\affiliation{%
  \institution{Simon Fraser University}
  \department{School of Computing Science}
  \city{Burnaby}
  \state{British Columbia}
  \country{Canada}
}
\email{borui_li@sfu.ca}

\author{Tingyu Zhang}
\affiliation{%
  \institution{Simon Fraser University}
  \department{School of Computing Science}
  \city{Burnaby}
  \state{British Columbia}
  \country{Canada}
}
\email{tza80@sfu.ca}

\author{Xing-Dong Yang}
\affiliation{%
  \institution{Simon Fraser University}
  \department{School of Computing Science}
  \city{Burnaby}
  \state{British Columbia}
  \country{Canada}
}
\email{xingdong_yang@sfu.ca}

\renewcommand{\shortauthors}{Yu et al.}

\begin{abstract}
    In this paper, we explore a novel approach that leverages retrofitting to create sensor-powered smart car cabins. We propose that ret\-ro\-fit\-ting offers a promising way to complement and extend the capabilities of built-in smart cabin sensors provided by car manufacturers. To understand how retrofitting solutions should be designed, we conducted a two-phase study. First, through semi-structured interviews with 18 participants, we examined challenges with built-in smart cabin sensors and identified opportunities where retrofitting could address these limitations. Second, through probe-based participatory design sessions with 15 participants, we identified user requirements and expectations for effective retrofit solutions. Based on our findings, we present a set of design recommendations to guide the future development of retrofit methods for smart car cabins.
\end{abstract}

\ccsdesc[500]{Human-centered computing~Empirical studies in HCI}

\keywords{smart car cabins, sensors, retrofitting}

\begin{teaserfigure}
  \includegraphics[width=\textwidth]{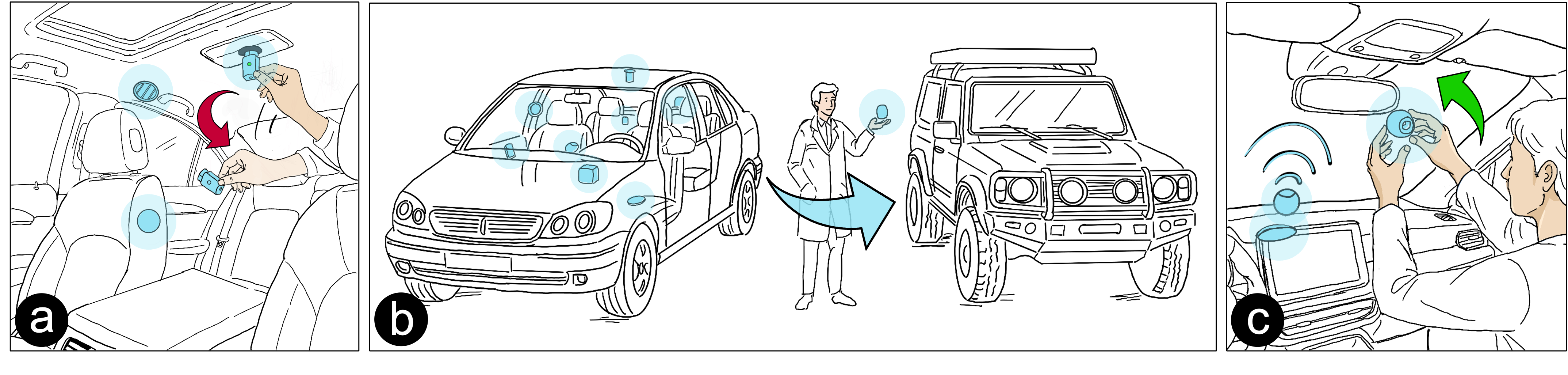}
  \caption{Illustration of the concept of retrofitting car cabins using user-selected sensors. A user can place various types of sensors, such as those monitoring air quality for climate control, gesture input, voice input, or biometric signals like fatigue, at different locations within the cabin to create customized smart cabin experiences. (a) The user can reconfigure the smart cabin experience by relocating or removing sensors. (b) The user can transfer their smart cabin experiences from one car to another. (c) The user installs sensors carried from another vehicle and sets them up in the new car, enabling familiar smart cabin experiences in the new vehicle.}
  \label{fig:figure_1}
  \Description{Illustration of the concept of retrofitting car cabins using user-selected sensors. A user can place various types of sensors, such as those monitoring air quality for climate control, gesture input, voice input, or biometric signals like fatigue, at different locations within the cabin to create customized smart cabin experiences. (a) The user can reconfigure the smart cabin experience by relocating or removing sensors. (b) The user can transfer their smart cabin experiences from one car to another. (c) The user installs sensors carried from another vehicle and sets them up in the new car, enabling familiar smart cabin experiences in the new vehicle.}
\end{teaserfigure}

\maketitle

\section{Introduction}
Modern automobiles are equipped with an increasingly diverse set of sensors. While many are designed to monitor the vehicle’s operational state, such as wheel speed sensors that detect slippage \cite{wheel_speed}, a growing number, particularly in newer models, capture user-related data to improve the in-cabin experience. Examples include humidity, temperature, and air-quality sensors that automatically adjust climate control \cite{air_quality}, mid-air gesture sensors for interacting with infotainment systems \cite{Gesture}, microphones for voice-based input \cite{voice_assistant}, and sensors that detect driver emotion or health status to help reduce accident risk \cite{Emotion}. These user-oriented sensors are typically installed by manufacturers during production and are often embedded discreetly, making them less visible to occupants \cite{hidden}. As a result, drivers and passengers are often unaware of their presence and cannot configure or select them when purchasing a vehicle. This lack of visibility and customizability can lead to sensor configurations that do not align with specific user needs or preferences (as shown in our study). 

Motivated by this gap, we explore retrofitting car interiors with aftermarket sensors. Retrofitting allows users to select, place, and configure sensors inside their cabin according to their own needs (Figure \ref{fig:figure_1}). While this approach is widely used in indoor environments, such as customizing smart homes and workspaces \cite{smart_home_sensor}, it remains largely unexplored in automotive contexts. Retrofitting has the potential to address challenges similar to those found indoors, such as customizing sensor-enabled user experiences \cite{smart_home_sensor}. However, the unique challenges and opportunities of introducing retrofitted sensors into cars are not well understood. In particular, key questions remain: (1) What challenges do drivers and passengers face with built-in sensors provided by car manufacturers? (2) What opportunities can retrofitting provide to address these issues? and (3) What design considerations are essential to ensure retrofitting effectively meets users’ needs?

In this paper, we investigate these questions through a probe-based participatory design approach, a method commonly used for uncovering opportunities for novel technologies \cite{probe}. Our study focuses on legacy vehicles that require a human driver or some level of human supervision behind the steering wheel. These vehicles represent the majority of automobiles in everyday use today. The study consisted of two phases. In the first phase, we conducted semi-structured interviews to understand the challenges drivers and passengers encounter when interacting with manufacturer-installed (or built-in) sensors. This phase revealed five key challenges: portability of smart cabin experiences, accommodating diverse user needs, customization, hardware upgrades, and repair. It also revealed opportunities for retrofitting as a promising strategy to address these challenges. In the second phase, we conducted probe-based participatory design sessions to collaboratively brainstorm and develop retrofit approaches. These sessions provided rich insights into user expectations and requirements for effective retrofit solutions. Building on these insights, we propose a set of design implications to inform the future development of retrofitting strategies for sensor-enabled smart car cabins.

This paper makes three key contributions. First, we present a novel approach that leverages retrofitting to create sensor-enabled smart car cabins. Second, based on interviews and design activities, we provide insights into users’ challenges with manufacturer-installed cabin sensors, identify opportunities where retrofitting can address these challenges, and outline user requirements and expectations for effective retrofit solutions. Third, we offer a set of design recommendations to inform the future development of retrofit methods in smart car cabins.
\section{Background and Related Work}
In this section, we review prior work on car sensors and existing approaches for retrofitting vehicles and indoor environments.

\subsection{Manufacturer-Installed In-Cabin Sensors}
The automotive industry has grown rapidly in recent years, driven largely by the rise of electric vehicles \cite{iea2025ev}. This growth has coincided with a significant expansion of the global automotive sensor market \cite{Publishing}, with many vehicles now equipped with over a hundred sensors \cite{New_Automotive_Sensors—A_Review}. Traditionally, manufacturer-installed sensors focus on the powertrain, chassis, and body \cite{Overview_of_automotive_sensors}, primarily supporting safety and functional tasks such as monitoring tire pressure \cite{pressure_sensor}, oil temperature \cite{micromachined_sensors}, airbag deployment \cite{accelerometers}, and exhaust emissions \cite{gas_sensor}. While essential for safety, these sensors offered limited interaction for drivers or passengers. The rise of advanced driver assistance systems (ADAS) and autonomous driving has shifted vehicles toward data-driven, perception-rich environments, incorporating cameras, radars, and LiDAR to enable lane keeping, adaptive cruise control, pedestrian detection, and collision avoidance \cite{Lidar, Driver_Assistance_Systems}.

At the same time, the concept of the smart cabin has emerged as a key area of automotive innovation. Unlike traditional vehicle sensors, which focus primarily on monitoring car operations, smart cabin systems enhance in-cabin experiences by enabling seamless interaction with vehicle computers, improving comfort, and delivering personalized services \cite{smart_cabin_sensor}. Examples include driver monitoring systems that detect drowsiness or distraction using cameras \cite{Fatigue_Detection}, ECG sensors that anticipate driving behavior \cite{ECG}, gesture sensors for mid-air control of infotainment systems \cite{Gesture}, and microphones that support voice interaction with virtual assistants \cite{voice_assistant}. 


\subsection{Creating Smart Environments via Retrofitting}
Sensor-powered smart environments, such as homes, workplaces, and vehicle cabins, are physical spaces enhanced with diverse computing and interaction technologies. These environments are designed to help people achieve both personal and professional goals \cite{NUGENT2014459, Cook_SmartE}. At their core, smart environments consist of an ecosystem of sensing and computing devices, including connected sensors \cite{Khalil_2014_WSN, Rayes2022}, actuators \cite{Rayes2022}, data processors \cite{cui_radio_2019}, and software systems \cite{Hejazi_platform_2018, Zhang_2021_IOTAI}.

A key step in realizing these environments is the effective deployment of sensors throughout physical spaces. A common research strategy retrofits everyday environments by attaching sensing devices directly onto surfaces in a ad-hoc manner \cite{Zhang_Vibrosight_2018,Zhang_Sozu_2019,Gupta_ElectriSense_2010,Laput_2017_Synthetic,harrison2008scratch,Iravantchi_SAWSense_2023}. This approach works across multiple sensing modalities, including electromagnetic (EM) \cite{NoiseMyCommand,SignalNoise,Laput_2017_Synthetic,Gupta_ElectriSense_2010,Zhang_2018_Wall}, vibration \cite{Iravantchi_SAWSense_2023,Zhang_Vibrosight_2018,harrison2008scratch,xiao2014toffee}, and light \cite{Li_Starlight_2016,zhang_flexible_2022,Laput_SurfaceSight_2019}. It enables applications from input detection to activity monitoring. For instance, ElectriSense \cite{Gupta_ElectriSense_2010} attaches sensors to power outlets to capture EM interference for appliance recognition, while StarLight \cite{Li_Starlight_2016} uses sparse floor photodiodes to detect light signals and reconstruct human posture. Vibration sensors have also been attached to floors, walls \cite{Zhang_Vibrosight_2018}, and on plumbing, gas, and HVAC systems \cite{Zhang_Sozu_2019} to infer user activities.

Retrofitting often requires technical knowledge in electronics and computing, and access to specialized tools, such as 3D printers, which are not widely available to everyday users. To lower these barriers, researchers have developed tools that help non-experts design electronic devices, sensors, and 3D enclosures, enabling their integration into physical objects or surfaces \cite{IoT,3d_print}. Beyond research, retrofitting has also become common in commercial products that convert conventional spaces into sensor-powered smart environments \cite{aqara2025sensor}. For instance, water-leak sensors can be placed on the floor to detect leaks in the home \cite{aeotec2025waterleak}. Door sensors can be easily attached to track whether doors are open or closed \cite{adt2025doorwindow}. Similarly, motion sensors mounted on ceilings can detect human presence to automate actions such as turning lights on or off \cite{xiaomi2025motionsensor2s}. 

\subsection{Retrofitting Car Cabins with Technology}
Retrofitting has been widely adopted in academic research as a means of exploring novel ideas \cite{Headlight_Leveling, fabric, seat}. For instance, prior work has augmented automotive textiles with electronic sensing layers, turning familiar elements such as headrest covers and seatbelt pads into interactive devices that enable new forms of tactile engagement \cite{fabric}. Similarly, researchers have embedded sensing capabilities into seat cushions to monitor physiological states, showcasing how everyday cabin components can be repurposed to support health-related applications \cite{seat}. Together, these studies highlight the flexibility and promise of retrofitting, but they provide limited insights into how such approaches might be designed with end users in mind.

Beyond academic research, retrofitting has also become increasingly significant in industry, where it is used to enhance interactivity, safety, and overall user experience in car cabins \cite{Temperature/Humidity_Sensor, enhauto2025, dasai2025, commaai2025, adas, Blind_Spot_Detection, parking}. For example, tangible interfaces such as physical buttons have been introduced to support more intuitive in-car interactions \cite{enhauto2025}, while social robots are being integrated into cabins to foster engaging experiences among passengers \cite{dasai2025}. Similarly, aftermarket systems such as advanced driver-assistance systems (ADAS) \cite{commaai2025, adas}, blind-spot detection \cite{Blind_Spot_Detection}, and parking assistance \cite{parking} demonstrate how retrofitting enables older vehicles to be equipped with features that were once only available in newer models.

Our literature review suggests that retrofitting provides a promising approach for integrating sensors into car interiors. While previous work has primarily emphasized technological advancements, relatively little research has explored retrofitting from the users’ perspective. In particular, there is a limited understanding of how retrofitting could help address challenges posed by manufacturer-installed sensors. Moreover, few studies have investigated how retrofitting strategies can be designed to better align with users’ needs, preferences, and expectations.

\section{Methods and Data Analysis}
\begin{figure}[t]
    \centering
    \includegraphics[width=\columnwidth]{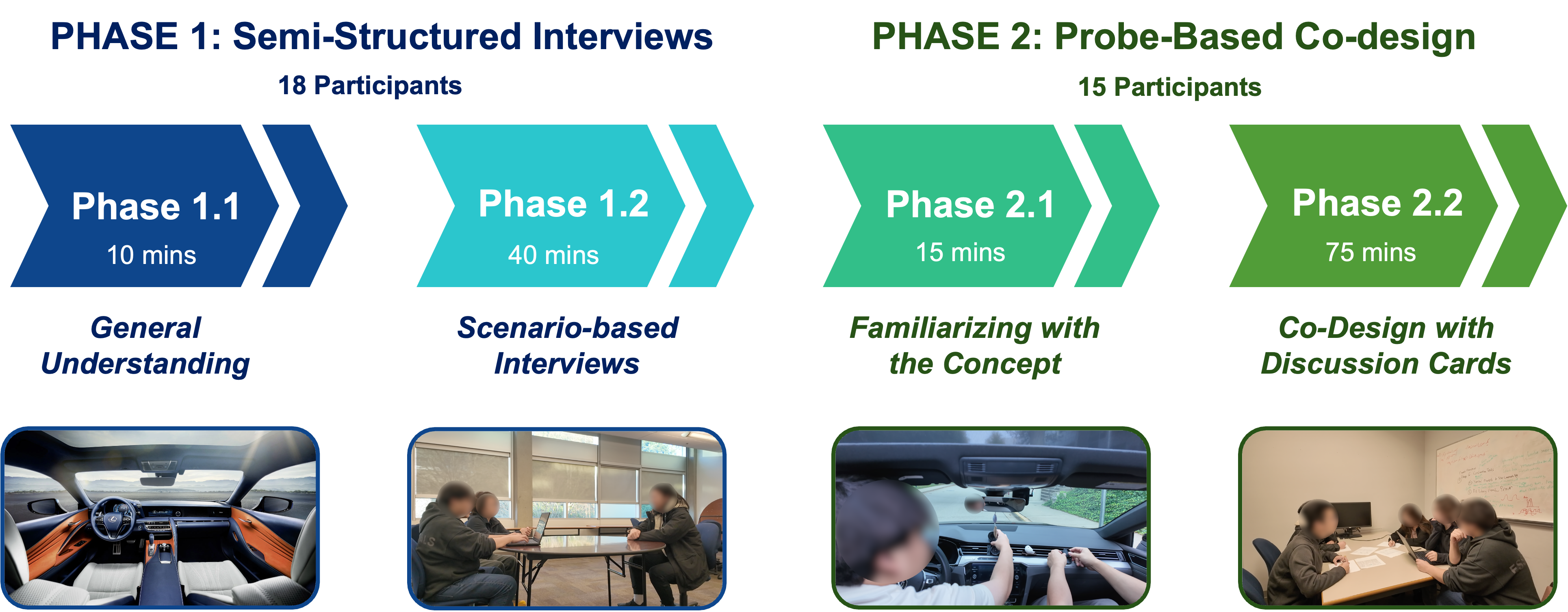}
    \caption{Overview of our two-phase study design. Phase 1 consisted of semi-structured interviews with 18 participants, including a 10-minute general understanding session (Phase 1.1) and a 40-minute scenario-based interview (Phase 1.2). Phase 2 involved probe-based co-design sessions with 15 participants, beginning with a 15-minute in-car familiarization activity (Phase 2.1) followed by a 75-minute discussion card based co-design session (Phase 2.2).}
    \label{fig: timeline}
    \Description{Overview of our two-phase study design. Phase 1 consisted of semi-structured interviews with 18 participants, including a 10-minute general understanding session (Phase 1.1) and a 40-minute scenario-based interview (Phase 1.2). Phase 2 involved probe-based co-design sessions with 15 participants, beginning with a 15-minute in-car familiarization activity (Phase 2.1) followed by a 75-minute discussion card based co-design session (Phase 2.2).}
\end{figure}

To address our research questions, we conducted a two-phase study (Figure \ref{fig: timeline}): (1) semi-structured interviews aimed at exploring the challenges drivers and passengers face when interacting with built-in smart cabin sensors, (2) Probe-based co-design sessions to explore retrofitting mechanisms for car cabins using aftermarket sensors. Insights from the first phase informed our design of the probes used in the second phase.

We collected audio recordings from both the semi-structured interviews (Phase 1) and the co-design sessions (Phase 2). The recordings were transcribed using an online transcription tool, and the resulting transcripts were carefully reviewed by the research team to correct recognition errors.

We then employed an open coding approach \cite{corbin2014basics}, allowing codes to emerge iteratively rather than relying on a predefined framework. Two authors independently conducted line-by-line coding of the transcripts to identify meaningful themes. An example of the code that emerged was “vehicles including sensors that users do not need”, which captures participants’ experiences of dissatisfaction with sensors built-in by car manufacturers that did not align with their preferences. Examples of the corresponding transcribe are (1) “More and more new cars rely on voice assistants to operate, but I’m not a fan of this way of interacting...” and “I am used to driving with one hand on the wheel, but now my new car detects whether I am holding it with both hands ... but it’s not a pressing need for me”.

During the open coding stage, we compared, refined, and grouped codes into clusters. We also applied axial coding \cite{vollstedt2019introduction} to explore relationships between codes. For example, “existing sensors fail to deliver desired experiences” and “vehicles include sensors that users do not need” were combined into the broader theme: “built-in sensors can hardly meet all user needs”. Codes and emerging themes were discussed and refined collaboratively until we reached consensus. This iterative approach helped ensure both consistency and depth in our analysis. We then selected representative quotes to highlight key findings, which were reviewed and verified by the two authors.

\section{Phase 1: Semi-Structured Interviews}
The first phase investigated the challenges users encounter when interacting with built-in smart cabin sensors. We conducted one-on-one interviews in a meeting room with 18 participants. Each session was moderated by one researcher while another observed and took notes. Interviews lasted approximately 50 minutes and was divided into two parts, described below.

\subsection{Study Design and Procedure}
The first phase introduced participants to the core functionalities of the smart cabin sensors, while the second phase explored the challenges users encounter when interacting with built-in sensors.

\hspace*{\parindent}\textbf{Part 1: General Understandings.}
We began our study by introducing participants to smart cabin sensors commonly integrated into modern vehicles. To facilitate this introduction, we used an interactive 3D visualization of a vehicle interior to illustrate the sensors and their typical locations within the cabin. We also explained their functionalities and representative application scenarios. Our overview focused on several widely used sensor categories, as shown below. Rather than presenting an exhaustive list, we emphasized commonly used sensors, guided by insights from the literature reviewed in the Related Work section.

\begin{itemize}
    \item Environmental sensors (e.g., humidity, temperature, and air-quality sensors) for automatic climate control.
    \item Gesture-recognition sensors (e.g., infrared or millimeter-wave radar) for interacting with infotainment systems or operating doors, windows, and sunroofs.
    \item Voice-input sensors (e.g., microphones) for voice command interaction.
    \item Driver monitoring sensors (e.g., millimeter-wave radar) for tracking vital signs, fatigue, and stress.
    \item Emotion and health detection sensors (e.g., RGB or infrared cameras).
    \item Posture and seating sensors (e.g., pressure sensors) for automatic seat adjustment and posture recognition.
\end{itemize}
These examples represent a broad set of sensors distributed across common cabin locations (e.g., seats, dashboard, ceiling) and application scenarios. This stage of the study was designed to calibrate participants, regardless of their prior experience with smart cabin technologies, so that they shared a common baseline understanding before proceeding to the next phase.

\textbf{Part 2: Scenario-based Interviews.}
The second part of our study followed an approach similar to \cite{OnlineBehavioralAdvertising,PrivacyPerceptionsSmartHome}, in which we introduced four scenarios to participants to help them envision different roles in the car (e.g., driver, passenger) and contextualize discussions around built-in smart cabin sensors. These scenarios reflected common use cases of vehicles and were inspired by prior research findings: (1) \textit{Rental Car (S1)}: Your job requires frequent business travel, and you often rent cars for transportation. You notice that the sensors installed in different vehicles vary considerably in both functionality and user experience \cite{rentalCar}. (2) \textit{Purchasing a Car (S2)}: You are considering buying your dream car and discover that it is equipped with multiple smart cabin sensors \cite{prieto2013exploration}. (3) \textit{Shared Family Car (S3)}: Your family shares an older vehicle, and you are planning to replace it with a newer model that offers various smart cabin sensors \cite{cycil2014designing}. (4) \textit{Multi-Purpose Car (S4)}: You are a part-time Uber driver who uses your personal car for both family needs and ride-hailing services. Your vehicle is equipped with smart cabin sensors \cite{ma2025analysis}.

The scenarios were designed to capture diverse vehicle-use contexts (e.g., driving unfamiliar vehicles, purchasing new vehicles, sharing with family members, or sharing with both family members and strangers) and user roles (e.g., primary driver, occasional driver, passenger, owner, and non-owner). During the discussions, participants were not restricted to the sensors introduced in Part 1. They were encouraged to reference smart cabin sensors they had previously encountered and to speculate on their potential use within the presented scenarios. Due to time constraints, each participant discussed two of the four scenarios. These scenarios were presented in a counterbalanced order using a Latin square design, to ensure that all scenarios were discussed an equal number of times across participants. After discussing the scenarios, participants demonstrated a clear understanding of both the benefits and potential issues of built-in smart cabin sensors. The conversation then shifted toward the challenges of built-in sensors. 

\subsection{Participants}
We recruited eighteen participants (11 males, 7 females) aged 23–30 (M = 27.1, SD = 2.5) (See Appendix A). All participants had at least three years of driving experience or comparable experience as passengers. Among them, one had an engineering background, seven had backgrounds in computer science, and the remaining participants came from non-engineering backgrounds, ranging from business to statistics. Participants also reported prior experience with smart cabin sensors, though most indicated familiarity with only a subset of the available sensors on the market.

\section{Phase 1 Findings: Challenges of build-in Smart Cabin Sensors}
Our results indicate several challenges that drivers and passengers encounter when interacting with built-in smart cabin sensors. Since prior work has explored privacy concerns in sensor-powered car cabins \cite{data_privacy2,koester2022perceived,bella2021car}, we focus our findings on non-privacy-related aspects. Specifically, we identify challenges related to portability of smart cabin experiences, accommodating diverse user needs, customization, hardware upgrades, and repair. Because these challenges arise independently of whether a vehicle is manually driven, supervised, or fully autonomous, they are likely to persist in fully autonomous vehicles as well. In the following sections, we elaborate on each challenge in detail.


\subsection{Built-in Sensors Limit the Portability of Familiar User Experiences Between Vehicles (C1)}
Many users enjoy smart cabin experiences and often wish to replicate them in another vehicle, such as a rental or work car (P2, P12, P15). However, variations in sensor availability across car brands and models make it difficult. Users must therefore adapt to a new interaction model when switching cars, which introduces a learning curve they find undesirable. Participants compared this to the transformable user experience common in mobile ecosystems, where app configurations and experiences remain consistent across phones, tablets, and laptops, but note that such consistency is largely absent in smart car cabins. For example, participants noted: 

\begin{quote}
    \textit{"I really like my car’s fatigue detection system. Lately, I’ve noticed that most rental cars don’t have this feature, which makes me feel less safe. I wish this kind of support could travel with me, no matter what car I’m in." (P12)}
\end{quote}

\begin{quote}
    \textit{"The touchscreen in my car detects when my hand approaches, and the control panel pops up automatically. I really like this and hope that this experience could extend to other vehicles I drive occasionally." (P15)}
\end{quote}

\subsection{Built-in Sensors Can Hardly Meet All User Needs (C2)}
Built-in cabin sensors typically follow a one-size-fits-all approach, designed to meet the needs of most users. However, our findings indicate that nearly all participants were dissatisfied with such sensors. Two common issues emerged: (1) built-in sensors often fail to deliver the full range of smart cabin experiences that users want, and (2) vehicles include sensors that users neither need nor notice. Consequently, users pay for features they do not value or use. For example, several participants (P5, P8, P11) described challenges transitioning from cars with traditional touch interfaces to newer models that prioritized voice input. Tasks previously simple became cumbersome, yet voice input remained a fixed feature that they could not opt out of when purchasing the vehicle. Other frustrations related to sensors that constrained personal driving habits but were bundled into the purchase. As P5 explained: 

\begin{quote}
  \textit{“I am used to driving with one hand on the wheel, but now my new car detects whether I am holding it with both hands. If not, it keeps notifying me. I understand it’s a part of safety features, but it’s not a pressing need for me.”}  
\end{quote}

Relatedly, many users are willing to explore new smart cabin experiences and interact with unfamiliar sensors. However, they expect a straightforward mechanism that allows them to opt out of any existing sensors at the hardware level (i.e., removing sensors they consider unnecessary). For example, participants stated:

\begin{quote}
    \textit{"I am open to trying new sensors, but the problem is that if I don’t like one, I can’t just return it like a smart home sensor. With cars, the only option is to return the entire vehicle, and that’s not what I want." (P14)}
\end{quote}

\subsection{Built-in Sensors Lack Customization (C3)}
As expected, built-in smart cabin sensors do not always meet personalized needs. In larger vehicles, such as minivans, voice and mid-air gesture sensors are typically positioned near the first two rows, since the back row is used less frequently. This placement reduces costs and is sufficient for most scenarios. However, for parents traveling with children, the back row is often in heavy use—either for car seats or to remain close to their kids. In these situations, passengers in the back may need to raise their voice to be heard by the microphone or stretch to reach the gesture sensors, as these sensors are not optimized for that row. Participants expressed interest in customizing sensor placement to be closer to them or in a location that supports easier interaction (P2, P9, P18). For example, participants shared:

\begin{quote}
    \textit{"Once I sat in the back of the van and had to shout just so people in the front could hear me. The van had a microphone that played passengers’ voices through the driver’s speaker, but it was too far from the last row. I still ended up yelling. I really wished I could just move the mic closer." (P2)}
\end{quote}

\begin{quote}
    \textit{"I’m left-handed, but the hand gesture sensor in my car was set up for right-handed users. I wish I could adjust its position to suit me better." (P9)}
\end{quote}

User preferences for sensor placement varied widely. For example, while many participants felt fixed camera positions were inadequate, but there was no consensus on an optimal placement. For example, P2 preferred ceiling-mounted cameras to capture the entire cabin, whereas P9 opposed ceiling placement, favoring positions less likely to capture their image due to privacy concerns. Preferences also shifted across use scenarios. P11 noted that during video conferences, they wanted the camera to face specific directions while avoiding others (e.g., preventing other passengers from appearing on screen). Current built-in cameras do not support such flexibility. 



\subsection{Built-in Sensors Are Hard to Upgrade (C4)} 
Participants perceived built-in smart cabin sensors as nearly impossible to upgrade, particularly for end users without technical expertise. Although upgrades might seem unnecessary, our results show that users often want to improve their cabin experience after seeing features available in newer models (P2, P12, P18). These improvements involve upgraded or new sensors, driven by the rapid pace of smart cabin innovation. In fast-growing markets, sensor-enabled features appear with almost every model release, sometimes within a year, and often rely on hardware that cannot be retrofitted. Unlike software, which can be updated over-the-air (OTA), hardware upgrades demand specialized skills, tools, space, money, and time. As one participant (P12) explained: 

\begin{quote}
    \textit{"I saw the new model of my car just came out with some great hardware and software updates. I know I’ll still get the software updates, but I was hoping the new hardware, like the depth-sensing camera, could be upgraded on my car too."}
\end{quote}

\subsection{Built-in Sensors Are Hard to Repair (C5)}
Another significant challenge with built-in smart cabin sensors is their limited repairability. To preserve a seamless appearance, these sensors are tightly integrated into the car interior, making them difficult for non-experts to access, repair, or replace. For instance, P10, who described themselves as “a pretty handy person” capable of performing basic car repairs such as replacing light bulbs, expressed frustration with the inaccessibility of embedded sensors. Participants also noted that when a single sensor fails, the entire car needs to be taken to a repair shop, which is both time-consuming and expensive. Drawing a comparison to their smart home devices, P6 explained:

\begin{quote}
    \textit{"If a door sensor at my home stops working, I can just remove it, inspect it, or replace it. But in my car, that’s impossible. I had to bring the entire car in for service. If my home sensor fails, I don’t take my whole house to Best Buy."}
\end{quote}

Another challenge arose when sensors do not perform as intended. Participants reported frustration with false triggers. Unlike at home, where users can easily unplug a smart speaker or turn off a security camera, vehicles offer no equivalent method to disable malfunctioning sensors. As P11 noted:

\begin{quote}
    \textit{"Voice commands kept triggering during my call, and I couldn’t just ‘unplug’ the sensor like I would at home."}
\end{quote}

\section{Retrofitting car cabins with aftermarket sensors}
We propose retrofitting as an alternative to pre-installing smart cabin sensors. Retrofitting has already proven effective in domains beyond the automotive context, particularly in the home, where it has been widely adopted to transform traditional spaces into sensor-enabled smart environments. 

We extend this concept to cars, proposing that similar benefits can be realized within the cabin. Specifically, car interiors can be retrofitted using sensor units, each designed for a dedicated function, such as a camera or a mid-air gesture sensor, or combining multiple capabilities, like an RGB camera, an IR camera, and a microphone, into a single unit. Inside the cabin, sensors can be attached to a variety of surfaces, including the ceiling, seat backs, or door panels. Once installed, they can be recognized and configured by the vehicle’s software, similar to device setup in smart home systems.

We identified several opportunities enabled by retrofitting that can help address challenges associated with built-in sensors. Ret\-ro\-fit\-ting allows users to select only the sensors they need and install them within the cabin, supporting personalized environments tailored to individual preferences (C2, C3). It also decouples smart cabin experiences from the vehicle itself (C2). Furthermore, retrofitting allows users to experiment with new sensors and experiences while retaining the flexibility to remove devices that do not meet their needs (C2). Users can also transfer their sensors across different vehicles, making smart cabin experiences portable (C1). Compared to built-in solutions, retrofitted sensors are more accessible, simplifying upgrades, maintenance, and repairs (C4, C5).


Although promising, prior work provides limited guidance on designing retrofitting approaches that align with users’ expectations and needs. Phase 2 of our study was designed to address this gap.

\section{PHASE 2: Probe-Based Co-design}
In this co-design study, we investigated how users envisioned the concept of retrofitting, their expectations for potential retrofit approaches, and the key requirements guiding these approaches. We recruited five groups of three participants each. Sessions were structured in two parts: first, participants sat inside a car to familiarize themselves with the idea of retrofitting the cabin with sensors. Second, they engaged in in-depth group discussions in a meeting room. Each session lasted approximately 90 minutes and was audio-recorded with participants’ consent for subsequent analysis. While our study primarily focused on legacy vehicles that require a human driver for operation or supervision, participants were free to extend their discussions to fully autonomous vehicles.

\subsection{Preparation}
We used a participatory design approach with discussion cards to immerse participants in potential retrofitting scenarios. Drawing on insights from Phase 1, we developed several use cases: (1) sensor migration, (2) sensor removal, and (3) multi-user cabin sharing to illustrate plausible challenges that may arise when retrofitting with aftermarket sensors, such as forgetting to bring a sensor, removing the wrong device, or managing trust in shared spaces. Each card presented one or two design directions proposed by the research team, covering both software- and hardware-oriented possibilities (see Appendix B). Rather than serving as prescriptive prototypes, the cards functioned as discussion prompts. Participants assessed the solutions, sharing preferences, concerns, and expectations. They were also encouraged to extend the ideas and propose alternatives, shifting the discussion from passive evaluation to active co-design. Overall, we found that the discussion cards provided participants with a tangible starting point that reduced ambiguity while leaving room for creative exploration.

\subsection{Study Design}
Our co-design sessions consisted of two parts. First, participants were introduced to the concept of retrofitting a car cabin with aftermarket sensors. Second, they explored potential retrofit designs. Each session was facilitated by two researchers: one moderated the activities, while the other observed, documented, and provided support.

\begin{figure}[t]
    \centering
    \includegraphics[width=\columnwidth]{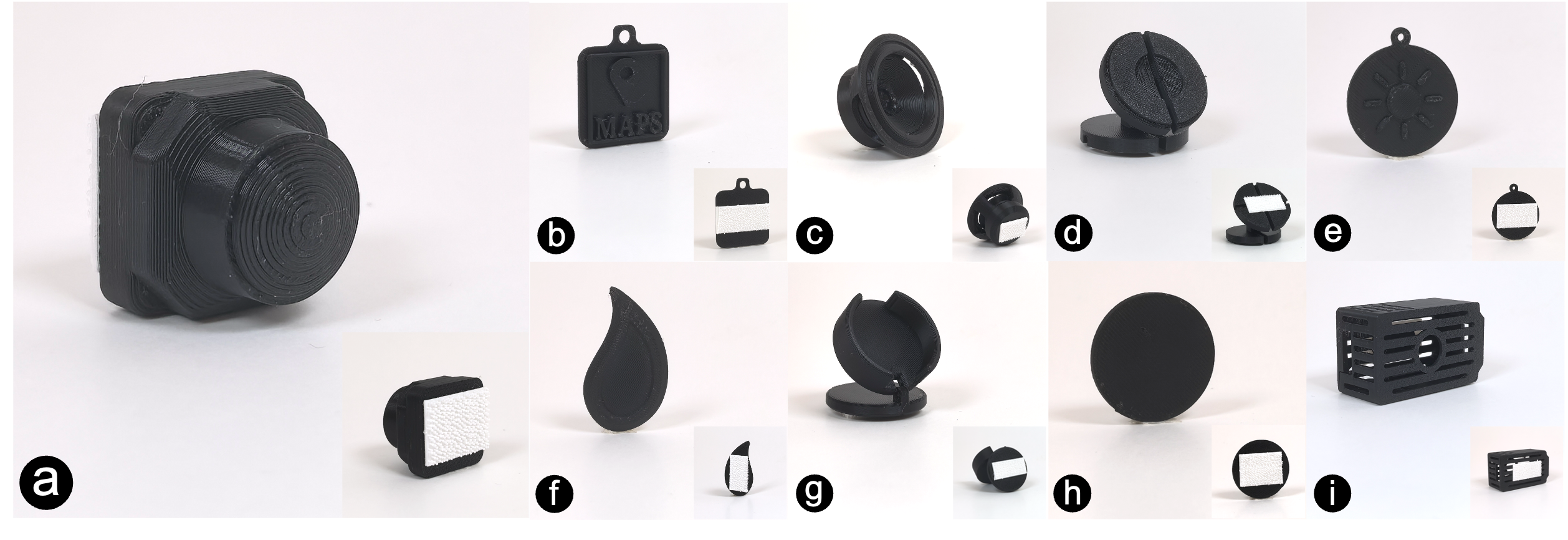}
    \caption{The 3D-printed sensor mockups used in our study. Each mockup includes Velcro on the back, allowing it to be attached to the Velcro strips placed at various locations inside the car interior. The set includes: (a) Camera (b) GPS (c) Microphone (d) Gesture Sensor (e) Light Sensor (f) Humidity Sensor (g) Millimeter-Wave Radar (h) Pressure Sensor (i) Temperature Sensor}
    \label{fig: sensors}
    \Description{The 3D-printed sensor mockups used in our study. Each mockup includes Velcro on the back, allowing it to be attached to the Velcro strips placed at various locations inside the car interior. The set includes: (a) Camera (b) GPS (c) Microphone (d) Gesture Sensor (e) Light Sensor (f) Humidity Sensor (g) Millimeter-Wave Radar (h) Pressure Sensor (i) Temperature Sensor}
\end{figure}

\textbf{Part 1: Familiarizing with the Concept of Retrofitting.}
The co-design session began with an introduction to the concept of retrofitting car cabins with aftermarket sensors. To help participants become familiar with the in-car context, we conducted a short hands-on activity in which they attached and detached mock-up sensors at various locations on the vehicle’s interior surfaces. This exercise simulated core retrofitting tasks, such as sensor installation and removal, and offered a simple, tangible experience of interacting with aftermarket sensors. Note that retrofitting workflows require careful design of both hardware (e.g., sensor attachment mechanisms and power delivery) and software interfaces (e.g., tools guiding users through the process). At this early stage of our investigation, we employed low-fidelity prototypes to focus on participants’ understanding of the overall concept rather than on specific implementation details. The session took place inside a 2019 Volkswagen Arteon four-door sedan. 

\begin{figure}[t]
    \centering
    \includegraphics[width=\columnwidth]{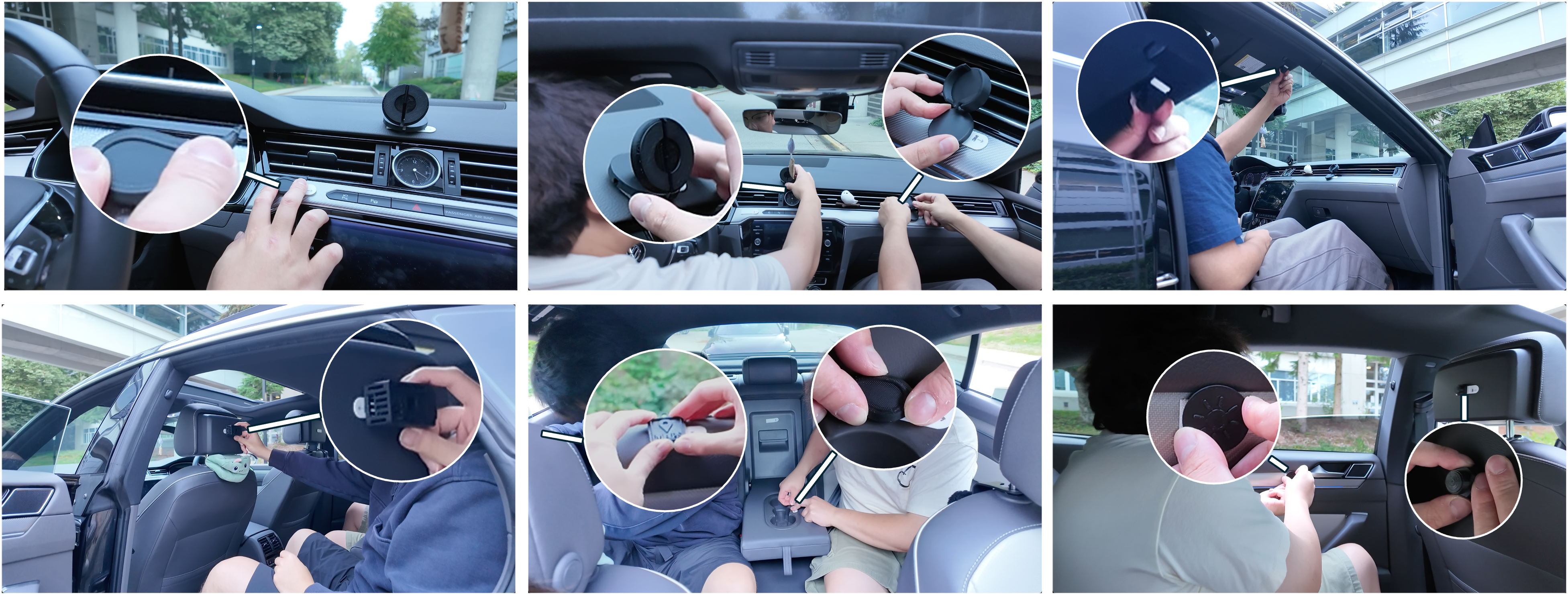}
    \caption{Participants interacted with the 3D-printed sensor mockups, experimenting with their placement in different locations inside a car.}
    \label{fig: install}
    \Description{Participants interacted with the 3D-printed sensor mockups, experimenting with their placement in different locations inside a car.}
\end{figure}

Our low-fidelity sensor prototypes (Figure \ref{fig: sensors}) were 3D-printed, allowing participants to physically handle them and get a sense of the potential form factors for different sensor types. The prototypes could be attached to car interior surfaces using Velcro, a simple mechanism chosen to minimize attention to technical implementation. During the study, participants were encouraged to attach the mockups to a range of locations, including the ceiling, door panels, and floor (Figure \ref{fig: install}).

Because realistic smart-cabin scenarios often involve multiple occupants, we invited multiple participants to be present in the cabin at the same time. This co-presence allowed them to observe how their sensor placement choices intersected with one another. At the start of the activity, participants were seated in the car with initial positions assigned randomly. They were then given the option to switch seats, allowing them to gain perspectives from different seating locations. Participants were encouraged to take notes throughout the activity but were instructed not to discuss the study topic with each other before the co-design session. They could explore the concept for as long as they wished, and the session concluded once all participants felt confident in their understanding. On average, each session lasted approximately 15 minutes.

\textbf{Part 2: Co-Design with Discussion Cards.}
This part of the study took place in a meeting room, where participants engaged in open discussions about cabin retrofitting using discussion cards as prompts. Before starting, we introduced the cards, clarified their content, and answered questions to ensure a shared understanding. The process was designed to be flexible: the cards served as conversation starters rather than prescriptive prototypes. Participants began by sharing their thoughts regarding the proposed solutions. Once initial opinions were voiced, we invited them to critique ideas, suggest refinements, or propose alternatives. Paper and pens were provided for sketching, although drawing was entirely optional.

\subsection{Participants}
Fifteen participants (7 females, 8 males) aged 20–30 (M = 26.1, SD = 3.03) were recruited for Phase 2 of the study (See Appendix A). Of these, eleven had also participated in Phase 1, while four were newly recruited. As in Phase 1, all newly recruited participants had at least three years of driving and/or passenger experience. They also reported prior experience with smart cabin sensing technologies. Among all participants, two had an engineering background, five had a computer science background, and the remaining participants had backgrounds spanning a range of non-engineering disciplines (e.g., Business and English literature). This diversity allowed us to capture perspectives from users with varying levels of technical expertise, providing insights into the broader population of car users.


\section{Phase 2 Findings: User Expectations of the Retrofitting Process}
Our participatory design sessions revealed five key themes capturing users’ requirements for retrofitting approaches: customizability, sensor installation, sensor removal, transferring the smart cabin experience, and multi-user scenarios. Our findings also shed light on some of the unique challenges and design requirements that arise when retrofitting AVs compared to legacy vehicles. We discuss them in this section.

\subsection{Participants' Response to the Concept}
Overall, participants responded very positively to the concept. They appreciated retrofitting as a novel approach for enabling smart cabin experiments. Many compared this approach to the challenges identified in the first study and expressed excitement about its potential to address issues arising from the current practice of manufacturers pre-installing sensors. Participants especially valued the freedom to choose their own sensors, decide where to position them, and determine when and how to use them. While some raised questions about whether the retrofitting process might be too complex or require specialized skills (discussed later), the overall response emphasized excitement about the benefits of this flexibility. For instance, several highlighted the value of decoupling the smart cabin experience from the vehicle, noting the possibility of extending smart cabin features, or even elements of their smart home, across different cars they drive or ride in. This enthusiasm was reflected in the new usage scenarios they envisioned after engaging with the discussion cards. For example, some participants (P10, P14) suggested carrying smart cabin sensors into hotel rooms to recreate cabin experiences while in their hotel.

Beyond participants’ enthusiasm, we identified several aspects of the retrofitting process that require thoughtful design to make cabin retrofitting a viable alternative to built-in sensors. We next present participants’ expectations and key user requirements to inform the design of future iterations of this technology.

\subsection{Customizability and Flexibility}
Participants’ design activities and discussions highlighted the value of giving users the freedom to select sensors that best match their preferences when retrofitting car cabins. Equally important is enabling users to position these sensors in locations that best align with their individual needs and intended use.

\textbf{Sensor location.}
Several participants valued the ability to position sensors based on individual needs. A driver, for example, may prefer to have a dedicated gesture sensor positioned near their hand by the center console or door, enabling them to control functions such as music without interference from other occupants. For passengers who are more comfortable sharing a gesture sensor, they could have the sensor placed in a more central and accessible location, such as the rear center armrest.

\begin{figure}[t]
    \centering
    \includegraphics[width=1\columnwidth]{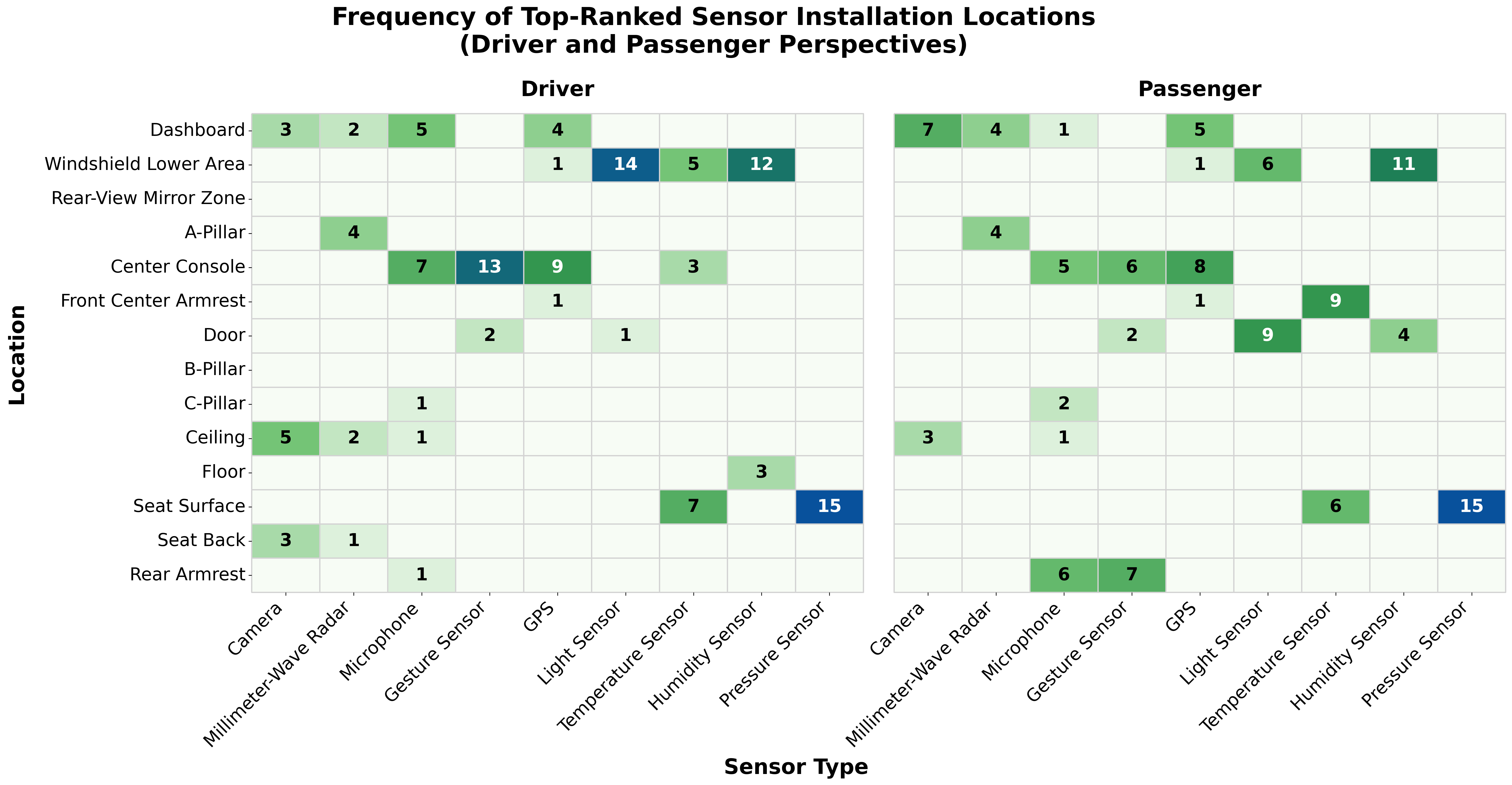}
    \caption{Illustration of the potential sensor installation locations identified by our participants, along with the frequency with which each location was ranked as the top choice for the sensors in this study. Results are shown separately from the perspective of the driver (left) and the passenger (right).}
    \label{fig: preferred_locations}
    \Description{Illustration of the potential sensor installation locations identified by our participants, along with the frequency with which each location was ranked as the top choice for the sensors in this study. Results are shown separately from the perspective of the driver (left) and the passenger (right).}
\end{figure}

During the study, participants identified a broad set of potential locations for sensor placement within the car cabin, including the dashboard, lower windshield area, center console, ceiling, areas near the rear-view mirror (including the mirror itself), the A-, B-, and C-pillars, floor, front and rear center armrests, seat surfaces (both seat and back), seat backs, and doors. Figure \ref{fig: preferred_locations} presents these locations and indicates how often participants selected each as their top choice for the sensors used in this study. Sensor placement preferences differed between driver and passenger roles. When role-playing as drivers, participants tended to favor locations within easy reach, such as the center console, rear-view mirror, and A-pillar. Interestingly, participants often associated themselves with vehicle owners in this role, which encouraged them to consider the broader cabin environment. As a result, they were more open to sensor placements farther from the driver’s seat, including ceiling areas near the rear seats or the backs of the front seats, particularly for sensors like cameras intended to monitor in-cabin activities.

In contrast, when role-playing as passengers, participants often viewed themselves as non-owners with limited access to sensor data. They demonstrated preferences strongly influenced by privacy considerations. Privacy-sensitive sensors, including cameras, were more acceptable when placed farther away or in locations that could be easily occluded. For example, rear-seat passengers preferred dashboard-mounted cameras that could be partially shielded by the front seats. Across both roles, safety remained a priority. Drivers showed limited interest in adding sensors around the windshield unless they supported safety-critical functions, such as millimeter-wave radar or ambient light sensing. Passengers, meanwhile, preferred sensors positioned close to them, such as on armrests or adjacent doors, to avoid distracting the driver. Finally, we note that these findings are based on a low-fidelity 3D-printed prototype. Participant preferences may change with higher-fidelity or interactive sensing systems, and thus these results should be interpreted with caution.

In reflecting on these wide range of preferences, participants noted that the “super unit” design proposed by the researchers (i.e., multiple sensors integrated into a single module) was less effective. Specifically, when a sensor within such a module is constrained to particular locations, the placement of the entire unit becomes limited, thereby limiting overall flexibility. As P16 explained:

\begin{quote}
    \textit{“What attracts me to retrofitting with aftermarket sensors is that I can freely adjust their positions. If they are designed as an integrated module, then the one sensor that requires a fixed spot will restrict everything else, and I lose the ability to move things around.” }
\end{quote}

\textbf{Individual sensors versus all-in-one units.}
In addition to valuing the flexibility of placement, participants also emphasized the benefits of using multiple individual sensors over a single all-in-one unit. They explained several practical advantages. For example, hardware upgrades could be performed more easily, as each sensor could be updated or replaced independently without requiring modifications to the entire system. In contrast, upgrading components within an integrated module would be considerably more difficult. Similarly, in the event of a malfunction, it would be much easier to replace a single sensor than to replace the entire integrated unit. Beyond these considerations, participants most strongly appreciated the freedom of customization that individual sensors afforded. By incorporating only the sensors they wished to install, users could configure the cabin according to their specific needs and preferences, ultimately supporting a more tailored user experience. P12 noted:

\begin{quote}
    \textit{"When manufacturers decide which sensors to integrate into a module, they are already defining the use scenarios for me, and often that includes sensors I do not even want. This design approach feels very similar to built-in systems, where my customization space is limited."}
\end{quote}

\subsection{Sensor Installation}
Participants most often directed their design efforts toward mechanisms that would facilitate easy sensor installation, reflecting their view that ease of setup is a central concern. They emphasized that aftermarket sensors should not require technical skills, special tools, or significant time to install. For sensors that require placement in specific locations to function properly, participants' design activities suggested the importance of clear guidance to help users understand these requirements. They also expressed a need for a simple mechanism that provides feedback on whether the sensors have been installed successfully and are operating as intended.

\begin{figure}[t]
    \centering
    \includegraphics[width=\columnwidth]{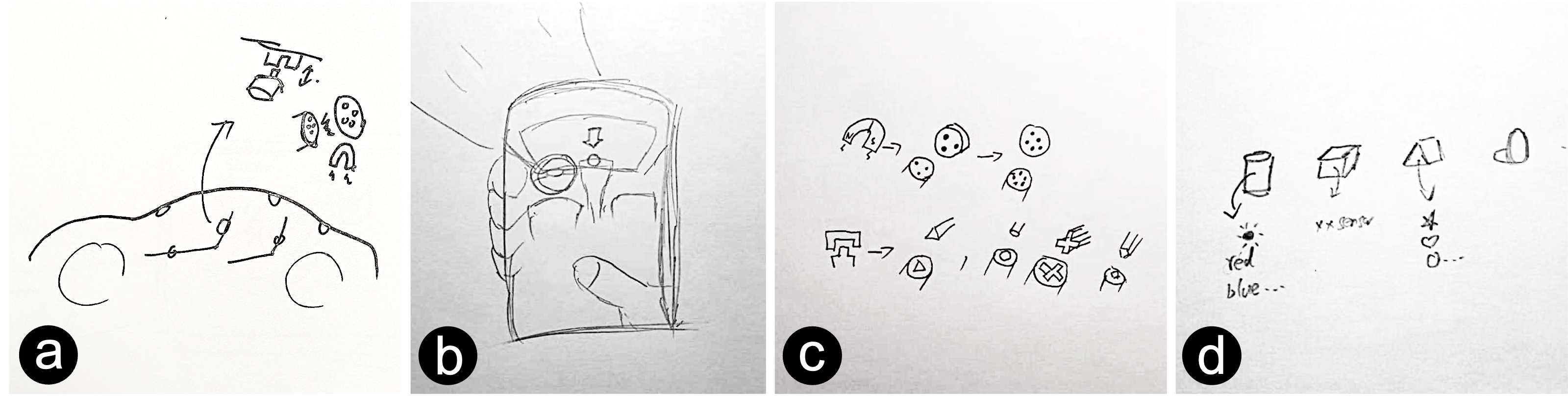}
    \caption{Participants’ sketches illustrating their envisioned approaches for retrofitting sensors into the car interior. (a) P19 proposed a modular plug-and-play design in which sensors can be attached to built-in slots using snap-in clips or magnetic mounts (b) P8 proposed using augmented reality to guide users in sensor placement. (c) P3 envisioned sensors with color coding, distinct icons to convey placement information. (d) The same participants envisioned sensors equipped with unique connectors (e.g., distinct magnet patterns or slot shapes) that guide users toward making the correct connection.}
    \label{fig: user_install}
    \Description{Participants’ sketches illustrating their envisioned approaches for retrofitting sensors into the car interior. (a) P19 proposed a modular plug-and-play design in which sensors can be attached to built-in slots using snap-in clips or magnetic mounts (b) P8 proposed using augmented reality to guide users in sensor placement. (c) P3 envisioned sensors with color coding, distinct icons to convey placement information. (d) The same participants envisioned sensors equipped with unique connectors (e.g., distinct magnet patterns or slot shapes) that guide users toward making the correct connection.}
\end{figure}

\textbf{Do I have the skills to install the sensors?}
Since aftermarket sensors require user installation, participants said they were open to this process if installation were simple and manageable. This was especially important for those eager to try new experiences and experiment with different sensors. Participants valued support for trial-and-error during setup, noting that without such support, they might be discouraged from adopting retrofit solutions in vehicles. Several participants compared the challenges to retrofit technologies in their homes, emphasizing that both contexts share similar barriers. Across nearly all groups, there was broad agreement that installation should not demand specialized knowledge or technical expertise. P19 elaborated on their design preferences as follows: 

\begin{quote}
    \textit{"I want the hardware to be modular and plug-and-play. It should hook right into the existing slots, like the ones already built into the ceiling, seat backs, or door panels, so I don’t have to deal with wiring or screwing."} (Figure \ref{fig: user_install})
\end{quote}

\textbf{Where should I install the sensors?}
A common concern among participants was that improper sensor placement could reduce sensor performance and lead to user frustration. Consequently, much of the design discussion focused on this issue. For sensors with specific requirements (e.g., air quality sensors, which should not be placed in areas with limited airflow), participants supported providing guidance or instructions to help users understand these requirements. Discussions also explored what forms of guidance might be most suitable, including digital manuals, augmented reality (AR) guidance (Figure: \ref{fig: user_install}), and similar approaches. Participants with prior experience using AR emphasized that such experiences should be carefully designed. They highlighted potential pitfalls, such as designs that require users to physically scan the entire car cabin to identify suitable sensor locations. Concerns about AR were also raised regarding accessibility and the learning curve for certain user groups, including older adults or people who are less technologically experienced, but who still drive or ride as passengers.

Participants also discussed designing mechanisms to help users avoid placing sensors in suboptimal locations. For example, their designs suggested that connectors deployed across interior surfaces and sensor bases could incorporate visual cues, such as color coding or distinct shapes, to convey placement information (Figure:\ref{fig: user_install}). As P3 noted:

\begin{quote}
    \textit{"Position-sensitive sensors could use specially designed connectors, like unique magnet patterns or distinct slot shapes that follow a consistent design language. This would make it harder to attach the sensors incorrectly, helping ensure they’re deployed correctly and intuitively."} (Figure \ref{fig: user_install})
\end{quote}

\textbf{Did I install the sensors correctly?}
Participants acknowledged that they often lacked confidence in their technical skills, especially in hardware installation. As a result, much of the discussion centered on how design strategies could help reduce this barrier. A common suggestion was creating testing environments that would allow users to verify their setup, identify potential errors, and ensure that the installation operates as intended. Within the cabin, the testing interface should be accessible to both drivers and passengers from any seat. Personal computing devices, such as smartphones, smartwatches, or emerging AR glasses, were brought up as complements to the car’s existing touchscreen interfaces as flexible testing platforms (P9, P13, P17).

\subsection{Sensor Removal}
Participants described several scenarios requiring sensor removal from the vehicle cabin. For example, they might want to reconfigure the sensors within the cabin, relocate them to different positions, physically disable sensors to stop their sensing functionality, or store them away when not in use. Note that this differs from typical smart home environments, where, once sensors are installed, users generally do not have the motivation or need to remove them. Participants emphasized several critical needs in these scenarios.

\begin{figure}[t]
    \centering
    \includegraphics[width=\columnwidth]{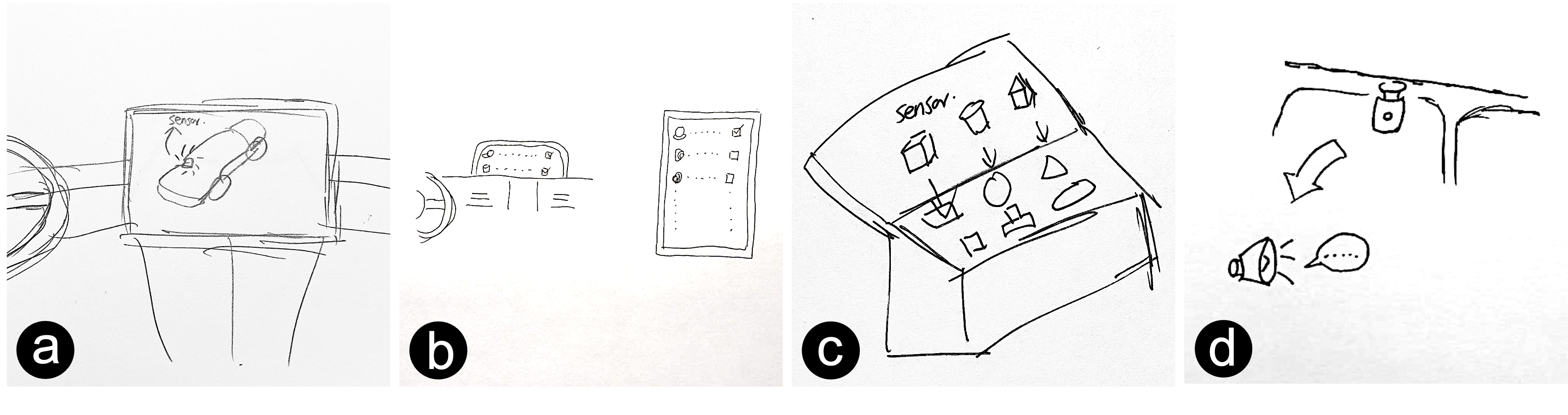}
    \caption{Participants’ sketches illustrating their ideas for sensor removal and storage. (a) P16 envisioned a car interface that visually guides users by indicating which sensors should be removed and showing their exact locations. (b) P9 proposed using a smartphone app to view sensor locations and related information inside the car. (c) P10 proposed a dedicated storage box designed with grooves shaped to fit different sensors, allowing unused sensors to be securely stored. (d) P16 suggested that the retrofitting system should inform users of the consequences after a sensor is disabled.}
    \label{fig: user_remove}
    \Description{Participants’ sketches illustrating their ideas for sensor removal and storage. (a) P16 envisioned a car interface that visually guides users by indicating which sensors should be removed and showing their exact locations. (b) P9 proposed using a smartphone app to view sensor locations and related information inside the car. (c) P10 proposed a dedicated storage box designed with grooves shaped to fit different sensors, allowing unused sensors to be securely stored. (d) P16 suggested that the retrofitting system should inform users of the consequences after a sensor is disabled.}
\end{figure}

\textbf{Which sensors should I remove?}
Before users can detach sensors, they first need to know which sensors should be detached. In a cabin retrofitted with multiple aftermarket sensors, users may recall the functions or experiences enabled by the sensors but often forget which specific sensor or sensors provide each experience. Therefore, a key user need is a mechanism that allows them to easily identify the target sensors they want to remove, understand their functionality, and locate them within the cabin. As suggested by P16:

\begin{quote}
    \textit{"I’d design the car system to include software that clearly shows which sensor needs to be removed. The software would list each sensor along with its function and corresponding location. Ideally, the sensor itself could also light up, making it even easier for users to find them."} (Figure \ref{fig: user_remove})
\end{quote}

In scenarios where the car cabin contains sensors from multiple people (e.g., during a road trip where friends each bring their own sensors), participants emphasized the importance of identifying the owner of each sensor, in addition to knowing which sensors should be detached. Since sensors can appear very similar, this feature helps prevent accidentally removing someone else’s sensor, which could disrupt that person’s experience. P9 suggested: 

\begin{quote}
    \textit{“If there is an app, it should include a filter that allows me to see the location of my sensors inside the car, or even my friends’ sensors, in case they need my help to remove them as well.” }(Figure:\ref{fig: user_remove})
\end{quote}

\textbf{What will happen if I detach this sensor?}
Even with tools to help users identify which sensors to remove, there is still a risk of disconnecting the wrong sensor. Design discussions focused on mitigating this risk by providing real-time feedback that clearly indicates which function has been disabled after a sensor is detached. Equally critical is a mechanism that communicates, before removal, the broader implications of detaching a particular sensor, ensuring users understand the potential impact. Smartphones have been suggested for this purpose. However, relying on an external device may not always be ideal in cars. In certain driving scenarios, users might need to detach a sensor while operating the vehicle, especially if the sensor is easily within reach (e.g., physically detaching a camera to protect their privacy). Removing the wrong sensor could disrupt an active function or cause unexpected behavior, introducing potential safety risks. In these cases, using an external device like a smartphone to check the consequences would be cumbersome, and, in the worst case, dangerous. As P20 noted: 

\begin{quote}
    \textit{"I like that I am informed about the consequences before removing a sensor, but I don’t want to depend on another device to get that information. What if my phone isn’t nearby? It should be provided directly on the sensor or through the car’s infotainment system."}
\end{quote}

Some participants preferred receiving feedback after removing a sensor rather than checking beforehand, as this was perceived to be more efficient. They reasoned that if they accidentally removed the wrong sensor, they could reattach it right away. As P16 noted:

\begin{quote}
    \textit{"I prefer that after I take out a sensor, the car system shows a message like ‘Feature X is interrupted because sensor Y was disconnected’. It’s more direct and efficient. Even if I remove the wrong one, I can just put it back." }(Figure:\ref{fig: user_remove})
\end{quote}

\textbf{Where do I store the sensors?}
After detaching the sensors from the cabin, participants emphasized the need for a dedicated storage solution, particularly when the sensors were not immediately needed. A common concern was the possibility of misplacing sensors or forgetting where they had been stored. Inspired by the charging cases of wireless earphones, participants imagined a purpose-built storage box tailored for retrofit sensors. For instance, P10 explained: 

\begin{quote}
    \textit{"The box would have grooves that match different sensors, so each one can only fit into its own slot. That way, I can immediately see if a sensor is missing. Ideally, the box would also ‘know’ which sensors are inside and which are not, with a companion app that lets me check this information."} (Figure \ref{fig: user_remove})
\end{quote}

\subsection{Migrating Experience}
In situations where users want to transfer their familiar sensor setup from one vehicle to another or reset a previous configuration, it is important to provide a way for them to easily identify the sensors necessary for the experience they wish to create. 

\textbf{How do I transfer my current experience to a new car?} 
Understanding which sensors are needed for the experience users want to replicate is only the first step. Car cabins vary widely in their structures and seating arrangements. For example, a coupe, a sedan, and a minivan each have very different layouts. As a result, a sensor configuration from one vehicle may not directly translate to another. Directly copying sensor placements could lead to the intended experience failing to transfer to a new car or require users to spend unnecessary time and effort identifying and resolving issues. Participants (P5, P16) suggested that when migrating sensor setups across vehicles, users should be guided on how to adjust sensor placement to accommodate the new interior and seating layout (potentially through a smartphone app). For instance, a system could indicate that while a sensor in the original car is located in one spot, the closest equivalent position in a rental car would be elsewhere. It is important to note that from the user’s perspective, their goal is to maintain a consistent experience across the vehicles they drive or ride in. Any additional effort required to migrate sensors could discourage them from adopting the retrofitting approach. 

\textbf{Can my experience be transferred selectively?}
Although migrating the full smart cabin experience to a new car might be ideal, participants found it more practical to transfer only the most frequently used features. Less frequently used features could be left behind, particularly in temporary migrations. This strategy reduces the effort of moving sensors between a temporary and a daily vehicle while still maintaining the core smart cabin experience. It is important to note that participants in our study often described the smart cabin in terms of overall feelings, such as comfort, efficiency, calmness, or a sporty vibe, rather than specific sensor interactions (P2, P18, P20). From their perspective, they are migrating their experience rather than the sensors themselves, highlighting a distinction between user-centered and technical viewpoints.

\begin{quote}
    \textit{"When I’m traveling, I don’t feel like I need to bring every sensor or smart cabin feature with me. As long as the essentials are there, I’d be happy. Just please don’t make me figure out which sensors I need to bring to the new car."}
\end{quote}

Ensuring a smooth transfer of sensors back to a user’s daily car is also important. Participants (P17, P22), who explored design strategies for this aspect of retrofitting, noted that users may be discouraged from adopting sensor migration if their current experience cannot be easily restored.

\begin{quote}
    \textit{"I want a way to save my installation plan. I don’t mind spending some time choosing and adjusting sensors and their locations the first time, but next time, when I want to restore that plan, I want a simple way to ‘copy’ and ‘paste’ my previous setup." (P17)}
\end{quote}

\begin{quote}
    \textit{"When I’m rushing to pick up a rental car, I don’t have time to redo the whole setup. I’d rather the system just load the key sensors I need immediately, and then I can slowly add the rest while I am already on the road." (P22)}
\end{quote}

\subsection{Multi-User Scenarios}
A unique usage scenario in vehicles could involve multiple users, each bringing their own sensors and collaboratively contributing to a shared smart cabin experience. This setup could enable novel and enhanced experiences for all occupants. However, if not carefully managed, it could also introduce potential confusion or conflicts of interest. 

Participants were excited about the scenarios where they could bring their own sensors and integrate them with those provided by other passengers in the cabin. Rather than operating in isolation, participants imagined these personal sensors working collaboratively to enable richer experiences. One scenario discussed involved a passenger providing a microphone to support voice commands, while another contributed a gesture or eye-tracking sensor to control the car’s infotainment system through gaze or hand movements. Together, these sensors could enable more natural interactions. For example, when a passenger issues the voice command “open the window”, the system might not know which window is being referred to. However, because people naturally accompany speech with gestures or gaze, the system could leverage input from gesture or eye-tracking sensors to more accurately infer the intended target. Participants emphasized the importance of mechanisms that make such possibilities visible and easy to enable, helping users understand how their sensors could interoperate with others’ to unlock new interactions (P16). They also highlighted the need for transparency about the consequences of combining sensors across passengers, so that users can make informed decisions about whether to participate in these new forms of interaction.

Sharing sensors among multiple cabin occupants might introduce privacy concerns. Occupants may not be aware of which sensors others have brought into the vehicle and what data is being collected. Participants discussed the scenario where one occupant might not realize that another has brought in a microphone, which could inadvertently compromise their privacy (P8, P18). Data ownership can also become more complex in these scenarios, as defining ownership is even more challenging than in traditional built-in sensor setups. Beyond privacy, shared sensors can lead to conflicting preferences. For instance, participants noted that one person might enjoy using voice commands while another might find them intrusive (P2, P5).

\subsection{Retrofitting in Autonomous Vehicles}
Although our co-design sessions focused on legacy vehicles with human drivers, participants frequently described opportunities for retrofitting fully autonomous vehicles (AVs), particularly those without steering wheels. They also articulated several ways in which AVs would introduce different design requirements. For example, participants viewed sensor placement as far more flexible in AVs than in legacy vehicles. In legacy vehicles, they were reluctant to place sensors on the windshield or rear-view mirror due to concerns about obstructing the driver (P17, P10), yet emphasized that such constraints would not apply in AVs (P2, P18). As P18 noted, \textit{“I would not be concerned about attaching sensors in these locations because no one would be driving"}. These insights suggest that preferred sensor locations in AVs warrant further investigation.

Participants also noted that the transition to AVs would expand the range of activities possible within the vehicle, particularly for those who would no longer need to drive. Freed from driving responsibilities, they anticipated using travel time for activities typically difficult to perform while driving, such as napping, watching videos, eating, working, or combining multiple activities during longer trips (P5, P8, P12). As a result, participants raised the need for sensors not traditionally associated with in-vehicle settings. For example, several participants emphasized the potential value of sensors that can track dietary behaviors, technologies more commonly found in indoor environments (P19, P21). They described scenarios such as bringing a lunchbox to eat during commutes, suggesting that diet-tracking sensors could meaningfully support these emerging routines. These findings suggest promising opportunities for extending smart cabin retrofitting to the ecosystems of indoor environments, which we further discuss in Section 9.4. 


Many participants anticipated a growing preference for ride-sharing with AVs over private car ownership, motivated by lower costs and greater convenience (P3, P8, P16, P19, P21). However, they noted that shared AV cabins were likely to offer only minimal smart features and would lack personalization. The discussion centered on how temporary retrofitting could address this gap, providing a degree of personalization for longer commutes (e.g., rides exceeding one hour (P3) during which activities such as eating or napping might take place). Participants emphasized that sensors should be easy to store and carry, and they highlighted the importance of enabling rapid testing and reconfiguration while the vehicle is in motion to adapt effectively to different cabin environments.

\section{Discussion}
Our studies reveal challenges with built-in sensors and highlight key design considerations for effective retrofitting. Based on these findings, we discuss the potential of retrofitting as an alternative approach, identify factors that may influence its adoption, and propose design guidelines to inform future development in this area.

\subsection{The Potential of Retrofitting as an Alternative to built-in Smart Cabin Sensors}
We propose retrofitting as a complement to built-in cabin sensors, which often limit drivers’ and passengers’ ability to personalize their smart cabin experiences. Our second study shows that user expectations for customization extend beyond sensing capabilities, which is commonly assumed to be the primary concern \cite{custom}. Participants emphasized preferences for sensor placement, which could inform technical design decisions, such as, whether sensors should operate independently or be integrated into a unified system. These preferences are also influenced by factors like ease of use, social acceptability, and privacy.

Retrofitting also promises portable cabin experiences across diverse contexts, from different vehicles to indoor and outdoor spaces. Prior research indicates that smart environment sensors are often developed in an ad hoc manner, optimized for a single target context, such as the home \cite{ding2011sensor}. While these designs perform well in their intended settings, our findings highlight the importance of creating sensors that support cross-environment applications. Although retrofitting is not primarily aimed at privacy, its features help mitigate some privacy concerns with built-in smart cabin sensors. Retrofitted sensors are often more visible to both owners and occupants, which prior work suggests can increase awareness and encourage strategies, like avoiding sensitive conversations near a speaker, that better align with their privacy preferences \cite{speaker_privacy}. 

In our first study, participants highlighted sensor upgrades and repairs as key considerations. Interestingly, these concerns were less prominent in the second study. One explanation is that the retrofitting approach naturally addresses these issues as sensors become more accessible. This also reflects the broader idea of decoupling the smart cabin experience from the car itself. Similar to how participants described augmenting homes with sensors to create smart homes, retrofitted sensors can be upgraded or repaired independently of the vehicle (e.g., without visiting a repair shop).

\subsection{Factors Affecting Retrofitting Adoption}
We observed several factors that could potentially influence user adoption of retrofitting. 

\subsubsection{Sensing Performance}
The promise of retrofitting lies in enabling sensors to perform as effectively when installed by users as they do when installed by manufacturers. Users may be hesitant to adopt retrofitting if the resulting smart cabin experience or sensor performance falls short of that offered by factory-installed systems. Unlike built-in sensors, retrofitted sensors must operate under dynamically changing conditions, such as suboptimal placement, while still delivering reliable and consistent performance.

\subsubsection{Usability}
Usability is another important factor influencing users’ adoption of retrofitting. Our study identified several important usability considerations, including the ease of installing and removing sensors, safe storage options, and the ease of transferring sensors between cars or other smart environments. These factors directly impact the overall usability of retrofitting and, consequently, influence users’ willingness to incorporate it into their smart cabin experience.

\subsubsection{Accessibility}
The accessibility of retrofitting is also important for user adoption. Cars are complex environments that people of diverse abilities and backgrounds engage with daily. The success of retrofitting depends on how easily drivers and passengers can adopt it, especially those who may have physical limitations, limited technical expertise, or constrained access to specialized tools or financial resources. Since vehicles are not fully accessible to all users \cite{milakis2020implications}, it is important to build on past experiences and prioritize accessibility as a core design consideration for effective retrofitting.

\subsubsection{Privacy}
While retrofitting offers opportunities to address some of the privacy concerns of built-in sensors, important challenges remain, particularly in shared smart car cabins with multiple occupants. In situations where different passengers bring their own sensors into the vehicle, questions arise around conflicting privacy preferences and data ownership. Prior research has demonstrated that smart environments that fail to account for diverse privacy expectations may limit user adoption \cite{smartE_privacy}. To ensure the successful adoption of retrofitting in automotive contexts, future work should carefully examine these issues. For instance, research could investigate how different roles, such as drivers versus passengers, or car owners versus non-owners, shape individuals’ privacy preferences and expectations regarding data ownership.

\subsection{Design Implications}
Our findings show that flexibility and customization are key reasons participants choose to retrofit. This aligns with prior work on user motivations for customizing software interfaces \cite{Banovic2012TriggeringTriggers, Griggio2019CustomizationsExpression} and smart-home sensing systems \cite{Jakobi2017SmartHomeLivingLab,Wozniak2023SmartHomeEcosystems}. By extending these insights to the automotive context, our work addresses the unique challenges vehicles present. For instance, installing smart cabin sensors often requires hardware skills, and retrofitting may introduce safety considerations while driving, challenges that are far less common in software customization or smart-home setups. Additionally, retrofitting can involve moving sensors between different vehicles, a scenario not typically encountered in home environments. In this section, we discuss the design implications derived from these unique factors to inform future efforts in retrofitting smart car cabins with sensors.

\subsubsection{Striving for Greater Customizability and Flexibility}
Building on the insight of customization, we recommend that retrofitting approaches prioritize empowering users to select sensors that meet their individual needs. Participants also valued customization beyond sensor functionality, emphasizing the importance of freedom in sensor placement and opportunities to tailor hardware design. Regarding placement, participants identified several factors that influenced their preferences. Beyond the basic requirement that sensors be installed in locations where they can reliably capture relevant events (e.g., gesture sensors should not be mounted beneath a seat), participants emphasized ergonomic aspects, such as ensuring accessibility from different seating positions, as well as concerns related to privacy, particularly in shared cabin scenarios. For example, they noted that sensors should be positioned to respect the comfort of passengers who may not wish to be monitored. To support flexibility, participants envisioned nearly all interior surfaces, including the windshield, ceiling, doors, floor, and seats, as potential locations for sensor placement, though some areas (such as the windshield) may introduce safety concerns for driving. While this vision may exceed current practical limits, it suggests an important avenue for future research to identify optimal regions for specific sensor types and how these choices align with user acceptance of different input modalities \cite{sensor_acceptance}.

Supporting this level of flexibility introduces several important technical challenges. First, sensors might need to be installable on non-flat or spatially constrained surfaces (e.g., door handles). Second, power delivery is limited, as the existing in-car infrastructure does not reach all interior locations. Wireless power transfer \cite{wirelessPower} presents a promising direction, although current techniques remain limited in capacity. From a hardware perspective, conventional design principles often prioritize structural integrity, compactness, and energy efficiency, typically favoring integrated, all-in-one solutions. In contrast, our findings suggest that users may prefer standalone sensors with distinct functionalities. While this preference may appear counterintuitive from an engineering standpoint, it provides valuable insights from a user-centered perspective. Participants highlighted the benefits of standalone sensors, noting their greater flexibility in placement and their ease of replacement, repair, and upgrading.

A promising future direction is the exploration of hybrid sensor systems, where modules can operate independently when distributed throughout the cabin, but can also be combined into a unified unit containing multiple sensor types. This approach could balance the flexibility valued in retrofitting and the usability challenges it may introduce. For instance, while participants noted that storing many detached individual sensors could become cumbersome, a modular unit containing several distinct sensors may mitigate this issue.

\subsubsection{Making Sensor Installation Easy for Users}
While the importance of easy and quick sensor installation may appear self-evident, it is particularly important in the automotive context. When installation is difficult, users are less likely to adopt retrofitting, as many may lack the time or technical expertise to manage complex setups. Designing for effortless installation also lowers the barrier for experimentation, enabling users to explore new sensor configurations and smart cabin experience. Based on our findings, we identify several strategies to facilitate installation.

First, installation should not require specialized skills or tools. Providing a plug-and-play workflow can substantially reduce barriers to adoption. For early deployment on flat, smooth surfaces commonly found in car cabins, such as windows, dashboards, and door panels, existing commercial attachment mechanisms can be leveraged, including vacuum suction pads \cite{suction_cup_wikipedia}. For non-flat or textured regions, such as fabric seats or the cabin ceiling, reusable adhesive tapes \cite{Tape3M9416} offer a practical alternative. Looking beyond commercially available solutions, emerging technologies such as electroadhesion \cite{8946902} present promising long-term opportunities. Although these techniques may require years of research before becoming practical for everyday use, they could ultimately broaden the range of viable installation sites. Regardless of the attachment method, all approaches should be evaluated carefully in the cabin environment. It is essential to evaluate how well each mechanism secures sensors against vibration and incidental contact, while still enabling effortless detachment when needed.

Second, users should receive guidance on suitable sensor locations, which is especially important when sensors have location constraints. Beyond textual instructions and illustrative figures, which could be shown on in-cabin displays, visual cues such as color-coded markers can be added directly to cabin surfaces to make placement and sensor orientation more discoverable. For applications that rely on mechanical interfaces (e.g., slots for power delivery), incorporating physical affordances, such as asymmetric slot geometries or directional magnetic poles, to further guide users toward correct and confident installation. 

Third, systems should provide immediate feedback that helps users confirm whether sensors are properly functioning right after installation. More importantly, this feedback should be accessible to all occupants, not just the installer, to promote transparency and shared awareness of sensor configurations within the vehicle. Upon completion of installation, sensors should provide visual or auditory cues to reassure users that the devices are active. More detailed status information, such as operational indicators, error messages, or debugging information, should then be accessible through in-cabin displays or the users’ smartphone app. For example, showing a camera’s video stream upon installation can quickly confirm that the device is properly installed.

\subsubsection{Facilitating Sensor Removal and Storage}
To support sensor removal, we propose three design considerations. First, systems should provide users with clear information about which sensors to remove, as our findings indicate that users do not always recognize which sensor corresponds to a specific function or experience. One potential approach is to enable the system software to display all active sensors within the cabin. This list could appear on the in-cabin displays or in a smartphone app. Through the app, users can select the function or experience they wish to disable (e.g., fatigue sensing), and the system can then highlight the corresponding sensors. Beyond software cues, physical sensors should also offer local visual or auditory feedback, such as flashing LEDs or short beeps, to help users identify the correct sensors and differentiate them from nearby sensors that should remain in place.

Second, the system should provide feedback and/or feedforward to help users understand the consequences of disabling or removing a sensor. For feedforward, when a target sensor is highlighted in the list, the system could communicate the effects of disabling it through animations or textual explanations. For example, the system should display a warning message on the in-cabin displays or the user’s smartphone app to indicate safety risks if a fatigue sensor is disabled. Once a sensor is disabled, the system should notify all cabin occupants, not just the user performing the action, ensuring that everyone remains informed about changes to the cabin’s sensor configuration.

Finally, providing dedicated storage ensures that detached sensors can be kept safe and organized. The storage could be implemented as a portable sensor case with individual slots for each sensor, allowing users to quickly identify which sensors are not in use. The case can be designed to detect the sensors placed within it, associate them with the corresponding vehicle, and communicate wirelessly with both the vehicle and the user’s smartphone app. This integration supports a real-time digital inventory, clearly indicating which sensors are stored and which are actively deployed within the vehicle cabin. Such a system facilitates the organized use of sensors and simplifies the transfer of sensors between cabins or other environments.

\subsubsection{Making Cabin Experience Easy to Transfer}
Our findings indicate that enabling a seamless transfer of the smart cabin experience to new cars or other environments offers promising opportunities. To support this transfer, we recommend three design considerations.

First, design for experience-centric transfer. It is crucial to focus on the experiences users want to replicate, rather than the underlying sensors. Participants in our study often described the smart cabin in terms of overall feelings, such as comfort, efficiency, calmness, or a sporty vibe, rather than specific sensor interactions. Migration support should therefore be experience-centric rather than sensor-centric. For example, users should describe the experience they wish to transfer without needing to specify individual sensors. Dialogues that facilitate this process should use clear, non-technical language.

Second, support selective transfer of experiences. Our study indicates that users may wish to selectively migrate certain aspects of their experience. For instance, they might choose to transfer only the most frequently used features. To support this, systems should provide mechanisms that help users identify their most-used features, such as visualizations of feature usage, and guide sensor selection based on which sensors are essential versus optional. These mechanisms should help users understand the minimum set of sensors needed to recreate their experience and the trade-offs involved when only a subset of sensors is available in a new car. Moreover, when users already know which experiences they want to transfer, the system should allow them to specify groups of experiences and provide tailored guidance throughout the migration process.

Third, support easy restoration of previous experiences. If users cannot easily restore their previous setup in their current car, they may be discouraged from adopting sensor migration. A naïve approach would be to repeat the same steps used when setting up a new car, such as restating their needs and repositioning sensors, which is redundant since this information has already been provided. For sensors with flexible placement, this could mean users would need to re-experiment with different locations if they forget the original configuration. To avoid such unnecessary effort, users should be supported with a straightforward way to save and retrieve their sensor configurations. 

\subsubsection{Making the Smart Cabin a Privacy-Respectful Environment}
Multi-user scenarios introduce unique challenges for retrofitting, particularly regarding privacy in shared cabins. Designing for these environments requires careful consideration, and we acknowledge that further research is needed to provide deeper insights into enabling privacy by design. One promising starting point is to enhance transparency by providing tools that clearly indicate which sensors are active, their functions, locations, sensing regions, the data they collect, and ownership. Users should also be empowered to control these sensors, potentially independent of ownership, allowing them to configure their functionality according to personal preferences.

\subsection{Transition to Fully Autonomous Vehicles}
Our findings center on legacy vehicles that still require some human involvement in their operation. While such vehicles continue to dominate today’s roads, AVs are expected to become increasingly common and will introduce new design challenges. Developing a deep understanding of user needs when retrofitting AV cabins will require further investigation, and our current work offers early insights into this emerging design space.

\subsubsection{Opportunities for Sensor Adjustment in Moving Vehicles}
In legacy vehicles, drivers must complete all retrofitting and sensor-adjustment tasks before operating the vehicle, as performing these actions while driving is unsafe. In contrast, AVs create opportunities for such tasks to be performed during the ride. This capability is particularly useful for adjusting sensor configurations, such as repositioning or recalibrating sensors, to optimize performance on the fly. Participants in our study noted that these adjustments, while sometimes necessary as in smart-home environments, are unsafe in conventional vehicles because they distract from driving. AVs, therefore, present new design opportunities for hardware and software systems that enable in-situ sensor configuration. Designers must consider unique constraints of the moving vehicle environment, including vehicle motion, limited space, and safety regulations that restrict user mobility.

\subsubsection{Rethinking Retrofitting for AV Cabins}
In legacy vehicles, the driver is often the primary user of the car and any added sensors. As our findings show, retrofit decisions often center on the driver’s needs. For instance, sensors designed to capture explicit input, such as voice commands, are commonly configured to optimize the driver’s experience or reduce distraction for safer driving. In AV cabins, the absence of a dedicated driver fundamentally changes this dynamic. Passengers with diverse needs may adjust or reconfigure cabin sensors to accommodate individual preferences, resulting in frequent, lightweight modifications. Consequently, the sensing environment may shift dynamically across different passenger groups. These changes can create confusion, especially for new passengers who may be unaware of the current sensing capabilities available in the cabin. While this situation is somewhat similar to smart home environments, where sensors are not typically tailored for a single inhabitant, AVs differ in that occupants change more frequently, and sensor configurations may be updated far more often than in homes or workplaces. These differences suggest the importance of designing AV retrofitting systems that prioritize transparency, awareness, and positive social interactions among passengers.

\subsubsection{Bridging Smart-Home and AV Ecosystems}
Our results indicate that the transition to AVs may enable new in-cabin activities that are uncommon in legacy vehicles, such as eating or napping during a trip. This shift suggests a need for novel sensors with new functionalities, which may also influence their preferred placement within the cabin. Many of these emerging activities are typically performed in indoor environments, and the sensors that support them are often found in homes or workplaces. Integrating these sensors into AVs suggests that the migration of smart-cabin experiences could extend beyond vehicle-to-vehicle interactions (Section 8.5), moving toward the incorporation of smart-home sensors within AVs. Designers and engineers may therefore need to consider how smart-home sensor hardware and software can also function effectively in AV cabin environments, and vice versa. This suggests the need for a carefully designed ecosystem that spans both indoor and in-vehicle settings.

\subsection{Expanding to Action-Driven Investigations }
Our work takes an initial step toward understanding how vehicle cabins can be retrofitted with aftermarket sensors. The findings and design implications from our research highlight many opportunities for further exploration, especially around action-driven questions of how these insights can guide the development of practical tools to support key retrofitting tasks, such as sensor installation, removal, and transferring experiences. They also raise questions about how effective such tools can be in real-world vehicle cabins, which often have limited space, diverse layouts, from traditional forward-facing seats to lounge-style interiors like those in Amazon Zoox \cite{ZooxRobotaxi2020}, and dynamic conditions that expose occupants to motion or vibration. For instance, installing sensors requires both mental and physical effort, as users need to interact with the sensors and in-cabin interfaces while processing real-time feedback and guidance. Designing effective tools for installation, therefore, involves addressing both technical and design challenges, including creating easy-to-attach and remove mechanisms, as well as intuitive solutions for feedback, guidance, sensor configuration, and debugging. 

Another key challenge is understanding the installation workflow. While the exact installation process needs further study, our initial findings suggest three key phases: (1) a pre-installation planning phase, where users prepare their retrofitting strategy before hands-on work; (2) an installation phase, in which users install sensors while receiving real-time guidance or feedback; and (3) a post-installation adaptation phase, where users iteratively reconfigure sensors based on interactions with the device, vehicle, cabin interior, and feedback from other stakeholders. This phased approach can also expand to other retrofitting tasks identified in our study, including sensor removal and storage, migration of cabin experiences, and configuring sensors to ensure privacy-respectful environments. In early stages, these investigations can leverage interaction probes, such as prototypes that respond to gesture or voice input, either in real cabins, like in our in-suit activities, or via VR simulations, to capture users’ reactions as they actively engage with retrofitting tasks.



\subsection{Limitations and Future Work}
Our study has several limitations that should be acknowledged. First, the co-design sessions relied on low-fidelity probes, including discussion cards and 3D-printed mockups. These probes were deliberately chosen to support creative exploration and lower barriers for participants to engage with speculative concepts. While they effectively facilitated idea generation and discussion, they could not fully replicate the functionality of working sensor systems. 
As a result, participants’ responses may differ from their behaviors in real-world environments. Future work could build on these findings by incorporating real sensors and devices, 
enabling a deeper understanding of realistic use patterns, installation practices, and potential adoption challenges.

Second, our analysis focused primarily on qualitative findings derived from interviews and co-design discussions. Although thematic analysis provided rich insights into challenges with built-in sensors as well as user needs and expectations for retrofitting, it does not quantify how widely these preferences are shared. Future work could adopt mixed-methods approaches, such as controlled experiments, to further validate and extend the findings from this study.

Third, our study examined retrofitting in legacy vehicles, offering insights into how retrofitting can be designed for the dominant vehicle type on today’s roads worldwide. While AVs are rapidly emerging, and some of our findings, such as expectations for customizability, may extend to that context, future work should carefully examine the specific design challenges and requirements of AVs. In particular, vehicles with alternative cabin configurations, such as forward-facing layouts found in Tesla’s Robotaxi \cite{TeslaRobotaxi2024} or the more open, lounge-style interiors of Amazon Zoox \cite{ZooxRobotaxi2020}, introduce unique considerations that merit further investigation.

Finally, our current work does not extend to designing or implementing tools to support retrofitting tasks, such as installing sensors. This was a deliberate choice, allowing us to focus on our core research questions: understanding the opportunities retrofitting offers to address challenges users face with built-in smart cabin sensors, and identifying design implications for various aspects of retrofitting. Future work can build on this foundation by exploring these critical tasks within the retrofit workflow. Key challenges include developing effective sensor attachment mechanisms, facilitating installation and removal, ensuring reliable power delivery across interior surfaces, maintaining compatibility with existing vehicle systems, and creating intuitive methods for sensor configuration and migration. Addressing these challenges offers an exciting opportunity for collaboration among HCI researchers, automotive engineers, and industry partners, and provides a pathway to further evaluate workflows and assess the effectiveness of retrofit approaches in supporting smart car cabin scenarios.










\section{Conclusion}
In this work, we investigate the potential of retrofitting car interiors with aftermarket sensors as a means to complement and extend manufacturer-installed sensing systems. Through a combination of semi-structured interviews and probe-based participatory design sessions, we uncovered key challenges drivers and passengers face with built-in sensors. Our findings suggest that retrofitting enables personalized in-cabin sensor configurations that better align with individual preferences. We also identified a set of user requirements that inform the design of effective retrofitting strategies. Based on these insights, we propose design recommendations to guide future implementations. This research demonstrates that retrofitting can not only complement existing systems but also offers a promising pathway toward realizing a broader ecosystem of smart environments. Beyond immediate retrofit applications, our study lays the groundwork for future exploration in modular sensor design, wireless power transfer in vehicles, and adaptive in-cabin interfaces. More broadly, this work contributes to the vision of smart ecosystems across diverse environments, including cars, homes, and outdoor spaces.

\bibliographystyle{ACM-Reference-Format}
\bibliography{references}

\clearpage 

\appendix

\section{Summary of Participants' Demographics}

\begin{table}[!h]
    \small
  \caption{Demographic of Participants}
  \Description{Demographic of Participants}
  \label{tab:demographic}
  \begin{tabular}{c c c c l}
    \toprule
    Participant & Gender & Age & Driving Experience (years) & Phases \\
    \midrule
    P1  & M   & 20-24 & 5 & Phase 1 \\
    P2  & M & 25-30 & 7 & Phase 1,2 \\
    P3  & M   & 25-30 & 10 & Phase 1,2 \\
    P4  & F & 25-30 & 5 & Phase 1 \\
    P5  & M   & 25-30 & 11 & Phase 1,2 \\
    P6  & M & 20-24 & 4 & Phase 1 \\
    P7  & F   & 25-30 & 9 & Phase 1 \\
    P8  & F & 25-30 & 10 & Phase 1,2 \\
    P9  & F   & 25-30 & 10 & Phase 1,2 \\
    P10 & M & 25-30 & 5 & Phase 1,2 \\
    P11 & M   & 25-30 & 11 & Phase 1 \\
    P12 & M & 25-30 & 11 & Phase 1,2 \\
    P13 & F   & 25-30 & 6 & Phase 1,2 \\
    P14 & M & 25-30 & 6 & Phase 1 \\
    P15 & M   & 25-30 & 7 & Phase 1 \\
    P16 & F & 25-30 & 11 & Phase 1,2 \\
    P17 & F   & 20-24 & 6 & Phase 1,2 \\
    P18 & M & 25-30 & 6 & Phase 1,2 \\
    P19 & M & 25-30 & 10 & Phase 2 \\
    P20 & M & 20-24 & 6 & Phase 2 \\
    P21 & F & 20-24 & 3 & Phase 2 \\
    P22 & F & 25-30 & 6 & Phase 2 \\
    
    \bottomrule
  \end{tabular}
\end{table}

\section{Discussion Card in Phase2: Probe-based Co-design}
\begin{figure}[H]
    \centering
    \includegraphics[width=0.9\columnwidth]{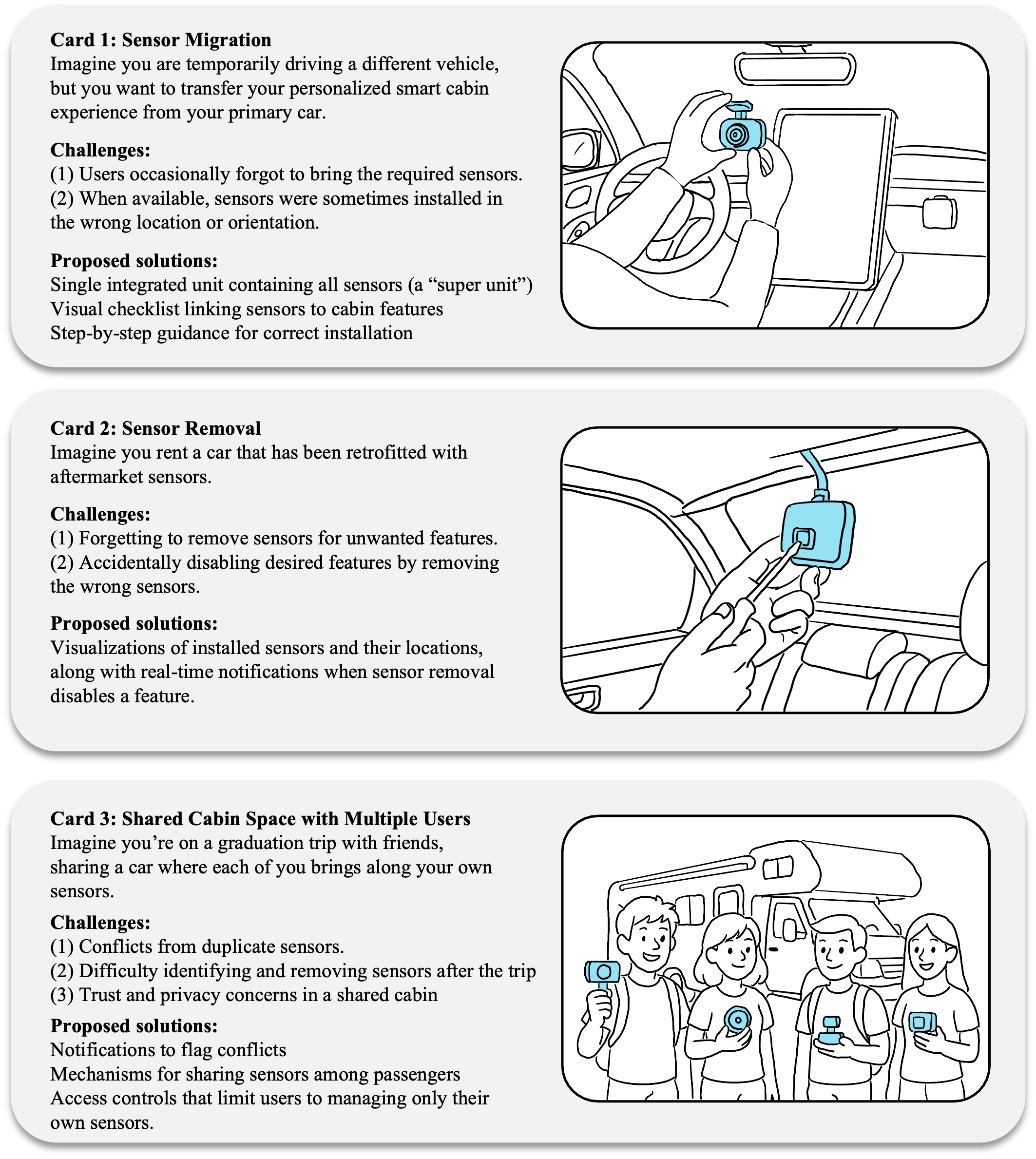}
    \caption{Discussion cards used during our co-design sessions. Each card presents a scenario along with the anticipated challenges and the research team’s proposed solutions. The three cards include: (1) Sensor Migration, (2) Sensor Removal, and (3) Shared Cabin Space with Multiple Users.
    }
    \label{fig: card}
    \Description{Discussion cards used during our co-design sessions. Each card presents a scenario along with the anticipated challenges and the research team’s proposed solutions. The three cards include: (1) Sensor Migration, (2) Sensor Removal, and (3) Shared Cabin Space with Multiple Users.}
\end{figure}

Figure \ref{fig: card} presents the discussion cards used in Phase 2 to scaffold co-design around smart cabin retrofitting scenarios. Each card illustrates a representative situation: (1) Sensor Migration, (2) Sensor Removal, and (3) Shared Cabin Space with Multiple Users. For each scenario, the card highlights two to three potential challenges intended to help participants reflect on the issues that may arise. The cards also outline several design directions proposed by the researchers, spanning both software and hardware possibilities. These cards were designed as prompts to stimulate discussion and inspire participant-driven ideas.    

\end{document}